\documentstyle[12pt]{article}

\begin{document}

%{\sf 

\begin{center}
\noindent
{\Large \bf Extended Thermodynamics to Einstein-Cartan Cosmology}\\[3mm]

by \\[0.3cm]

{\sl L.C. Garcia de Andrade\footnote{Electronic mail: garcia@symbcomp.uerj.br}}
\\[0.3cm]

Departamento de F\'{\i}sica Te\'orica\\[-3mm]
Universidade do Estado do Rio de Janeiro -- UERJ\\[-3mm]
Cep 20550, Rio de Janeiro, RJ, Brasil   

\vspace{2cm}
{\bf Abstract}
\end{center}
\paragraph*{}
The thermodynamics is extented to spacetimes with spin-torsion density.Impplications to Einstein-Cartan-de Sitter inflationary phases are discussed.A relation between the spin-torsion density,entropy and temperature is presented.A lower limit for the radius of the Universe may be obtained from the spin-torsion density and the Planck lenght.

\newpage

\section{Introduction}
\paragraph*{}

Riemannian extensions to the second law of thermodynamics to  General Relativity have been considered by H.P. de Oliveira and D.Pavon \cite{1,2}
in recent years.Since the spin dynamics seems to be fundamental on the mechanism of statistical mechanics and themodynamics and since spin dynamics is reasonably well explained in the context of Einstein-Cartan theory of gravity,the recent interest of thermodynamics on the investigation of cosmic background microwave radiation (CBMR) distribution in the Universe motivated us to investigate here the thermodynamics in the context of some inflationary models.We first solve the Gibbs equation in the realm of Einstein-Cartan gravity for a closed cosmological model and solve it for the radiation and inflationary eras.The next step is to solve the Einstein-Cartan cosmological equations for the de Sitter inflationary case.In this example we obtain a type of Ising model like a relation between the spin-torsion density and the absolute temperature and entropy density,producing a coherent model for thermodynamics in cosmology where the temperature and entropy increases with the spin-torsion density.Early attempts to include thermodynamics into the scope of Einstein-Cartan gravity were given by R.Amorim and L.L.Smalley \cite{3,4} by making use of a Lagrangean treatment.Nevertheless no attempts had been made by those authors to apply their theories to Cosmological problems.The COBE data \cite{5} of the CMBR to compute the spin-torsion fluctuation since it is shown that the temperature fluctuation measured by COBE is proportional to the spin-torsion fluctuation. 
\paragraph*{}
\section{Thermodynamics in Closed Einstein-Cartan Cosmological Models}
Let us start by defining the entropy density $s$ as \cite{5}
\begin{equation}
s=\frac{{\gamma}{\rho}}{T}
\label{1}
\end{equation}   
From the conservation law in EC gravity
\begin{equation}
\dot{\rho}+3H({\rho}+p)=2{\pi}(\dot{{\sigma}^{2}}+3H{\sigma}^{2})
\label{2}
\end{equation}
where we have addopted $G=1$ units and the equation of state is given by
\begin{equation}
p=({\gamma}-1){\rho}
\label{3}
\end{equation}
where ${\gamma}$ is the adiabatic index.Substitution of equation (\ref{3}) into (\ref{2}) yields
\begin{equation}
\dot{\rho}+3H{\gamma}{\rho}=2{\pi}(\dot{{\sigma}^{2}}+3H{\sigma}^{2})
\label{4}
\end{equation}
From the conservation law (\ref{3}) one obtains
\begin{equation}
\frac{d({\rho}'R^{3})}{dt}=-3p'R^{3}
\label{5}
\end{equation} 
where 
\begin{equation}
{\rho}'={\rho}-2{\pi}{\sigma}^{2}
\label{6}
\end{equation}
and
\begin{equation}
p'=p-2{\pi}{\sigma}^{2}
\label{7}
\end{equation}
are respectively the effective energy density and pressure including the spin-torsion density  
\begin{equation}
{\sigma}^{2}=<S_{ij}S^{ij}>=\frac{n^{2}h^{2}}{R^{6}}
\label{8}
\end{equation}
where $R(t)$ is the cosmic scale factor and $S_{ij}$ is the spin-torsion density tensor.Here ${i,j=0,1,2,3}$ and $n$ is the nucleon number and $h$ is the Planck constant.From the Gibbs relation
\begin{equation}
T\frac{dS}{dt}=\frac{dE}{dt}+p\frac{d({2{\pi}^{2}R^{3}})}{dt}
\label{9}
\end{equation}
Here $2{\pi}^{2}R^{3}$ is the volume of the spatially open $k=+1$ cosmological model \cite{5} and  
\begin{equation}
S=sR^{3}
\label{10}
\end{equation}
where S is the total entropy and s is the entropy density.By making $p'=0$ into expression(\ref{5}) one obtains 
\begin{equation}
T\frac{d({sR^{3}})}{dt}=0
\label{11}
\end{equation}
Substitution of (\ref{1}) into (\ref{11}) yields 
\begin{equation}
\frac{d{\frac{({\gamma}{\rho}R^{3})}{T}}}{dt}=0
\label{12}
\end{equation}
Expansion of expression (\ref{12}) yields
\begin{equation}
\frac{\dot{\gamma}{\rho}R^{3}}{T}-\frac{\dot{T}}{T}{\gamma}{\rho}R^{3}+\frac{{\gamma}\dot{\rho}R^{3}}{T}+3\frac{R^{2}\dot{R}{\gamma}{\rho}}{T}=0
\label{13} 
\end{equation}
Substitution of the conservation law into expression (\ref{13}) yields
\begin{equation}
\frac{\dot{\gamma}}{\gamma}-\frac{\dot{T}}{T}-3H({\gamma}-1)+2{\pi}[\dot{{\sigma}^{2}}+3H{\sigma}^{2}]=0
\label{14}
\end{equation}
Taking the factor ${\gamma}={\gamma}_{0}=constant$,after some algebra we obtain
\begin{equation}
lnT=3(1-{\gamma}_{0})R(t)+\frac{1}{\pi}\int{\frac{d{{S}_{0}}^{2}}{R^{2}}}
\label{15}
\end{equation}
where here $S^{2}_{0}=2{\pi}^{2}{\sigma}^{2}R^{3}$
is the total spin-torsion quantity.Expression (\ref{15}) yields
\begin{equation}
T(t)=e^{-3({\gamma}_{0}-1)R(t)+\frac{1}{\pi}\int{\frac{d{{{S}_{0}}^{2}}}{R^{2}}}}
\label{16}
\end{equation}
Substitution of the relation 
\begin{equation}
dS^{2}_{0}=-\frac{6{\pi}n^{2}h^{2}dR}{R^{4}}
\label{17}
\end{equation}
which upon integration and substitution on the last term on the RHS of equation(\ref{16}) yields
\begin{equation}
T(t)=e^{-3({\gamma}_{0}-1)R(t)+\frac{30{\pi}n^{2}h^{2}}{R^{7}}}
\label{18}
\end{equation}
which shows that the spatially closed Friedmann-Robertson-Walker model with spin-torsion density would cool down as the spin-torsion density increases, even for values of this density as big as the one corresponding to th nucleon number of the Universe $n=10^{80}$ nucleons.Another important physical result of the above result is that in the torsion free case as shown by Starkovich and Cooperstock \cite{5} during the inflation era ${\gamma}<\frac{2}{3}$ ,$\dot{T}>0$ while during the radiation era ${\gamma}=\frac{4}{3}$ ,$\dot{T}<0$ and $T$ reaches a maximum at the beginning of the standard model era.Let us now examine what happens in the more general case of spacetimes with spin-torsion density.In this case we have
\begin{equation}
\frac{\dot{T}}{T}=2{\pi}[\dot{{\sigma}^{2}}+3H{\sigma}^{2}]+3H(1-{\gamma})
\label{19}
\end{equation}
On the radiation era this equation reduces to
\begin{equation}
\frac{\dot{T}}{T}=2{\pi}[\dot{{\sigma}^{2}}+3H{\sigma}^{2}]-H
\label{20}
\end{equation}
Therefore the condition $\dot{T}<0$ together with equation (\ref{20}) implies
\begin{equation}
[\dot{{\sigma}^{2}}+3H{\sigma}^{2}]<\frac{3H}{2{\pi}}
\label{21}
\end{equation}
Substituting the value of the spin-torsion density in terms of the cosmic scale distance $R(t)$ yields ${R_{min}}^{3}>nh=nc^{3}l^{2}_{P}$ where $l_{P}$ is the Planck lenght.Therefore the spin-torsion density places a lower limit in the radius of the Universe avoiding the cosmological Singularity!As early as 1978 Kuchowicz \cite{6} has obtained similar results withour recurring to a thermodynamical model.Using Friedmann-like models and spin=spin interactions he was able to obtain for example for an Euclidean model a  minimum radius $R_{min}=\frac{{S_{0}}^{2}}{4D}$ where D is an integration constant.

\section{Thermodynamics on de Sitter Inflationary Phase in Einstein-Cartan Gravity }
\setcounter{equation}{0}

\paragraph*{} 
In this section I shall  consider an example of a de Sitter Cosmological model with spin-torsion density and thermodynamics.From the EC equations for this case one obtains
\begin{equation}
H^{2}=\frac{8{\pi}}{3}({\rho}-2{\pi}{{\sigma}^{2}})
\label{22}
\end{equation}
The other equation is given by
\begin{equation}
H^{2}=-\frac{4{\pi}}{3}(({\rho}+3p)-4{\pi}{{\sigma}^{2}})
\label{23}
\end{equation}
Since $p=({\gamma}-1){\rho}$ from equation (\ref{23}) one obtains
\begin{equation}
H^{2}=-\frac{4{\pi}}{3}((3{\gamma}-2){\rho}-4{\pi}{{\sigma}^{2}})
\label{24}
\end{equation}
By equating equations (\ref{22}) and (\ref{24}) one obtains
\begin{equation}
{\sigma}^{2}=\frac{{\gamma}{\rho}}{2{\pi}}
\label{25}
\end{equation}
Substitution of expression (\ref{1}) into (\ref{25}) yields
\begin{equation}
{\sigma}^{2}=\frac{sT}{2{\pi}}
\label{26}
\end{equation}
This expression seems to be analogous to the Ising model for the behaviour of spins on a ferromagnet.Besides it shows that the increase on the spin-torsion density represents an increase on the temperature and entropy density.Expressing the spin-torsion density in terms of the cosmic scale $R(t)$ allow us to obtain an expression between the temperature total entropy and the cosmic scale factor given by 
\begin{equation}
T=\frac{2{\pi}n^{2}h^{2}}{SR^{3}}
\label{27}
\end{equation} 
From the standard cosmology we know that 
\begin{equation}
T=\frac{1}{R}
\label{28}
\end{equation}
which yields
\begin{equation}
\frac{{\delta}T}{T}=-\frac{{\delta}R}{R}
\label{29}
\end{equation}
From the expression for ${\sigma}^{2}$ one obtains
\begin{equation}
{\delta}{\sigma}^{2}=-6\frac{(nh)^{2}{\delta}R}{R^{7}}
\label{30}
\end{equation}
which finally implies from the COBE data that
\begin{equation}
\frac{{\delta}T}{T}=\frac{1}{6}\frac{{\delta}{\sigma}^{2}}{{\sigma}^{2}}=10^{-5}
\label{31}
\end{equation}
which allow us to determine the spin-torsion density fluctuation from the COBE data \cite{7}.To better compare our thermodynamics to their ferromagnetic analogous Amorim-Smalley Lagrangean could be used to obtain a Hamiltonian in terms of the spin-torsion density and to compare the results with the Ising model.This can be done elsewhere.
\vspace{1cm}
\noindent
{\large\bf \underline{Acknowledgements}}: I am very much indebt to Prof.L.L.Smalley and H.P.de Oliveira for their interest in our work.Financial support from CNPq (Brazilian Government Agency) and UERJ are gratefully acknowledged.

\end{document}